\documentclass[pra,floatfix,twocolumn,superscriptaddress]{revtex4}
\usepackage{times,graphicx,bbm,amsmath,amssymb}
\usepackage[caption = false]{subfig}
\usepackage{epsfig,color}
\usepackage{hyperref}

\def\Tr{\hbox{Tr}} 
\usepackage{pifont}

\newcommand{\ket}[1]{\vert#1\rangle}
\newcommand{\bra}[1]{\langle#1\vert}

\begin{document}

\title{Metrology with Unknown Detectors}

\author{Matteo Altorio} 
\affiliation{Dipartimento di Matematica e Fisica, Universit\`a degli Studi Roma Tre, Via della Vasca Navale 84, 00146, Rome, Italy}
\author{Marco G. Genoni} 
\affiliation{Department of Physics \& Astronomy, University College London, 
Gower Street, London WC1E 6BT, United Kingdom}
\author{Fabrizia Somma} 
\affiliation{Dipartimento di Scienze, Universit\`a degli Studi Roma Tre, Via della Vasca Navale 84, 00146, Rome, Italy}
\author{Marco Barbieri} 
\affiliation{Dipartimento di Scienze, Universit\`a degli Studi Roma Tre, Via della Vasca Navale 84, 00146, Rome, Italy}

\begin{abstract}
The best possible precision is one of the key figures in metrology, but this is established by the exact response of the detection apparatus, which is often unknown. There exist techniques for detector characterisation, that have been introduced in the context of quantum technologies, but apply as well for ordinary classical coherence; these techniques, though, rely on intense data processing. Here we show that one can make use of the simpler approach of data fitting patterns in order to obtain an estimate of the Cram\'er-Rao bound allowed by an unknown detector, and present applications in polarimetry. Further, we show how this formalism provide a useful calculation tool in an estimation problem involving a continuous-variable quantum state, i.e. a quantum harmonic oscillator.
\end{abstract}

\maketitle

\section{Introduction}

With the introduction of quantum metrology, a clear framework has been established  for understanding metrological protocols as constituted of three steps: the preparation of the probe, its evolution through the interaction with the target system, and finally the extraction of information from a measurement~\cite{Giov04, Giov06, Pari08}. When targeting the best precision, these steps can not be taken as independent, in that the probe needs optimisation on the specific interaction, and, in turn, the  measurement needs to take into account both preparation and interaction to achieve the best precision. This description is effective not only when using quantum objects, but it can be adopted for designing {\it quantum-inspired} optimal protocols in the  classical domain, such as for position sensing~\cite{Kalt08, Dixo09}, and for polarimetry~\cite{Host08, Pfei11}. However, fundamental limits in measurements can be hard to achieve, since further limitations can originate in imperfect, or even faulty behaviours of the set-up. 

A possible solution for mitigating the effect of non-ideal measurements consists in adapting the design of the probe~\cite{Dorn09,Datt11,Knys11,Geno11,Geno12}. This normally requires a trustworthy characterisation of the measurement device which can be achieved by means of detector tomography~\cite{Lund09}, {\it i.e.} by reconstructing the action of the measurement device given the outcomes from a quorum of input preparations. In the quantum description, the knowledge on the device is cast as matrices linking the input probability amplitudes to that of each outcome~\cite{Hels76}; once these matrices are known, the optimal probe state can be found as the one minimising the variance of a valid estimator.

In this paper we discuss a different, more direct approach to the design of optimal probes when dealing with detectors departing from an idealised description, by making use of the {\it data fitting patterns} associated to the device~\cite{Reac10,Coop14}.  

\section{Quantum estimation}
\textcolor{black}{Let us consider the following quantum estimation problem for a set of unknown parameters $\vec\phi$. The quantum system is typically prepared in a pure probe state $|\psi_0\rangle$ and the interaction with the sample is described by a quantum completely positive map $\mathcal{E}_{\vec\phi}$. A measurement $\mathcal{M}$, formally corresponding to a positive operator valued measure (POVM), is then performed on the output state $\varrho_{\vec\phi}=\mathcal{E}_{\vec\phi}(|\psi_0\rangle\langle\psi_0|)$, and delivers an estimator $\vec\phi'$, {\em i.e.} a mapping from the experimental data to the parameter space. The data set should be sufficiently large to ensure that the estimator is unbiased, {\it i.e.} that the expectations value $\mathbbm{E}[\vec\phi']=\vec\phi$. The uncertainties on the individual parameters, as well as their correlations are captured by the covariance matrix $\underline{\Sigma}{=}\{\Sigma(\phi_i,\phi_j)\}_{i,j}$~\cite{Pari08}, whose elements are given by $\Sigma(\phi_i,\phi_j)=\mathbbm{E}[(\phi'_i-\phi_i)(\phi'_j-\phi_j)]$. }
 
\textcolor{black}{
The quantum and classical Cram\'er-Rao bounds establish that upon the realisation of $M$ experiments, the covariance matrix is limited as
\begin{equation}
\underline{\Sigma}\geq \frac{\underline{F}^{-1}(\vec\phi)}{M} \geq \frac{\underline{H}^{-1}(\vec\phi)}{M}
\end{equation}
where we have introduced respectively the quantum Fisher information (QFI) matrix $\underline{H}$ and the (classical) Fisher information (FI) matrix $\underline{F}$. The former is a property of the output state only $\varrho_{\vec\phi}$, and its elements can be evaluated as $H_{i,j}(\vec\phi){=}\Tr(\rho_{\vec\phi}\,[L_i,L_j]_+)$ where the symmetric logarithmic derivative operators $L_i$ are defined as 
\begin{equation}
\partial_{\phi_i}\rho_{\vec\phi}=\frac{1}{2}\left(L_i \rho_{\vec\phi}+\rho_{\vec\phi}L_i\right):=[\rho_{\vec\phi},L_i]_+ \:,
\end{equation}
and $[A,B]_+$ denotes the anti-commutator.
On the other hand the FI matrix is associated to a given measurement $\mathcal{M}$, and its elements are defined through the conditional probabilities $p(m|\vec\phi)$ of observing the outcome $m$ given the values $\vec\phi$:
\begin{equation}
\label{classfisher}
	F_{i,j}(\phi) = \sum_m \frac{\left(\partial_{\phi_i}p(m|\vec\phi)\right)\left(\partial_{\phi_j}p(m|\phi)\right)}{p(m|\vec\phi)}.
\end{equation}
While a measurement saturating the quantum Cram\'er-Rao bound always exists for single-parameter estimation, this is not always attained in the general case, and trade-offs have to be established, depending on the interest of each parameter.}

\section{Data fitting patterns} 

Once the optimal performance of the experiment has been designed \textcolor{black}{in terms of probe state $|\psi_0\rangle$ and measurement $\mathcal{M}$}, then the closest measurement setup has to be implemented. This will unavoidably depart from the ideal case, resulting in a decrease of information; further, if particular symmetries of the measurement are affected, undesired correlations between the estimates of the parameters can be introduced. These effects can be reduced if one seconds the actual measurement, and utilises instead a different probe state $\ket{\xi_0}$. In order to obtain a recipe for preparing $\ket{\xi_0}$, one needs full information on the action of the detector, which would demand a tomography of the measurement apparatus. This demands an optimisation routine on the experimental output probabilities $q_\alpha(m)$, collected when sending into the device a set of fiducial states $\ket{\alpha}$; the routine finds the closest well-defined measurement matrices $\Pi_m$ associated to each outcome~\cite{Lund09,Brid12,Zhan12,Gran15}. Then, using the expression for the classical Fisher information, Eq.~\eqref{classfisher}, the optimal state is found using Born's rule $p(m|\vec\phi){=}\Tr\left(\mathcal{E}_{\vec\phi} (\ket{\xi_0}\bra{\xi_0})\,\Pi_m\right)$. 

Here we discuss how the need for the first routine can be circumvented, and a more direct approach to data analysis is possible. The idea is that through the data fitting patterns of the detector, one obtains an explicit expression for the Fisher information, which can be used to get the optimal state. Data fitting patterns (DFPs) are defined as the output probabilities $q_\alpha(m)$ for an over-complete set $\{\ket{\alpha}\}$, by which any state $\rho$ can be written as $\rho{=}\sum_\alpha c_\alpha \ket{\alpha}\bra{\alpha}$~\cite{Reac10}. If the detection scheme is informationally complete, then any state can be reconstructed using the fact that its outcome probabilities are, by linearity $p(m){=}\sum_\alpha c_\alpha q_\alpha(m)$; state tomography is indeed the original aim for the introduction of this formalism. Linearity also ensures that the Fisher information Eq.~\eqref{classfisher} admits a decomposition
\begin{equation}
\label{dfpfisher}
F_{i,j}(\phi) = \sum_m \frac{\left(\sum_\alpha \partial_{\phi_i}C_\alpha(\vec\phi)q_\alpha(m)\right)\left(\sum_\alpha \partial_{\phi_j}C_\alpha(\vec\phi)q_\alpha(m)\right)}{\sum_\alpha C_\alpha(\vec\phi)q_\alpha(m)}.
\end{equation}
where $\{C_\alpha(\vec\phi)\}$ are the coefficients of the state $\mathcal{E}_{\vec\phi} (\ket{\xi_o}\bra{\xi_0})$. Therefore, the Fisher information can be optimised based on the knowledge of the data fitting patterns only, and it does not require reconstructing the elements $\Pi_m$, which could be computationally demanding. It should be noticed that the advantage is purely in the post-processing stage, since the requirements on the cardinality of the over-complete set are the same for detector tomography as for the DFPs~\cite{Reac10,Coop14}. 

The application of the DFP method in this case differs from the original proposal in that it can be applied to informationally incomplete measurements, {\it i.e.} measurements which are not sufficient for a complete tomographic reconstruction. This extension is possible because, in the general case, the parameters we need to estimate are a limited set with respect to the complete set defining a quantum state univocally in the relevant Hilbert space. This philosophy has been applied to the reconstruction of photon statistics with pseudo-number resolving detectors~\cite{silb}.


\section{Experimental example 1: single-qubit projective measurement.} 

We start with a simple example of phase-estimation with single qubits. Although we cast it in the language of quantum information, this problem is fully equivalent to optimal phase estimation in classical polarimetry. Our aim is to measure a small phase $\phi{\sim}0$ using an approximate projective measurement, implemented by a half-wave plate set at the angle $\theta$, and a polarising beam-splitter. The DFPs are collected by measuring the output intensities using the over-complete set of six states corresponding to the eigenstates of the three Pauli operators; for polarisation qubits these are given by four linear  ($H$orizontal, $V$ertical, $D$iagonal and $A$ntidiagonal), and two circular ($R$ight- and $L$eft-handed) polarisations. Since the problem involves a single parameter, optimisation simply consists in maximising the value $F\overset{\text{def}}{=} F_{1,1}$ of the Fisher information \eqref{dfpfisher}, by choosing the appropriate coefficients $C_\alpha(\phi)$, with $\alpha{\in}\{H,V,D,A,R,L\}$. 

\begin{figure}[t]
\includegraphics[viewport = 40 50 780 530, clip, width=\columnwidth]{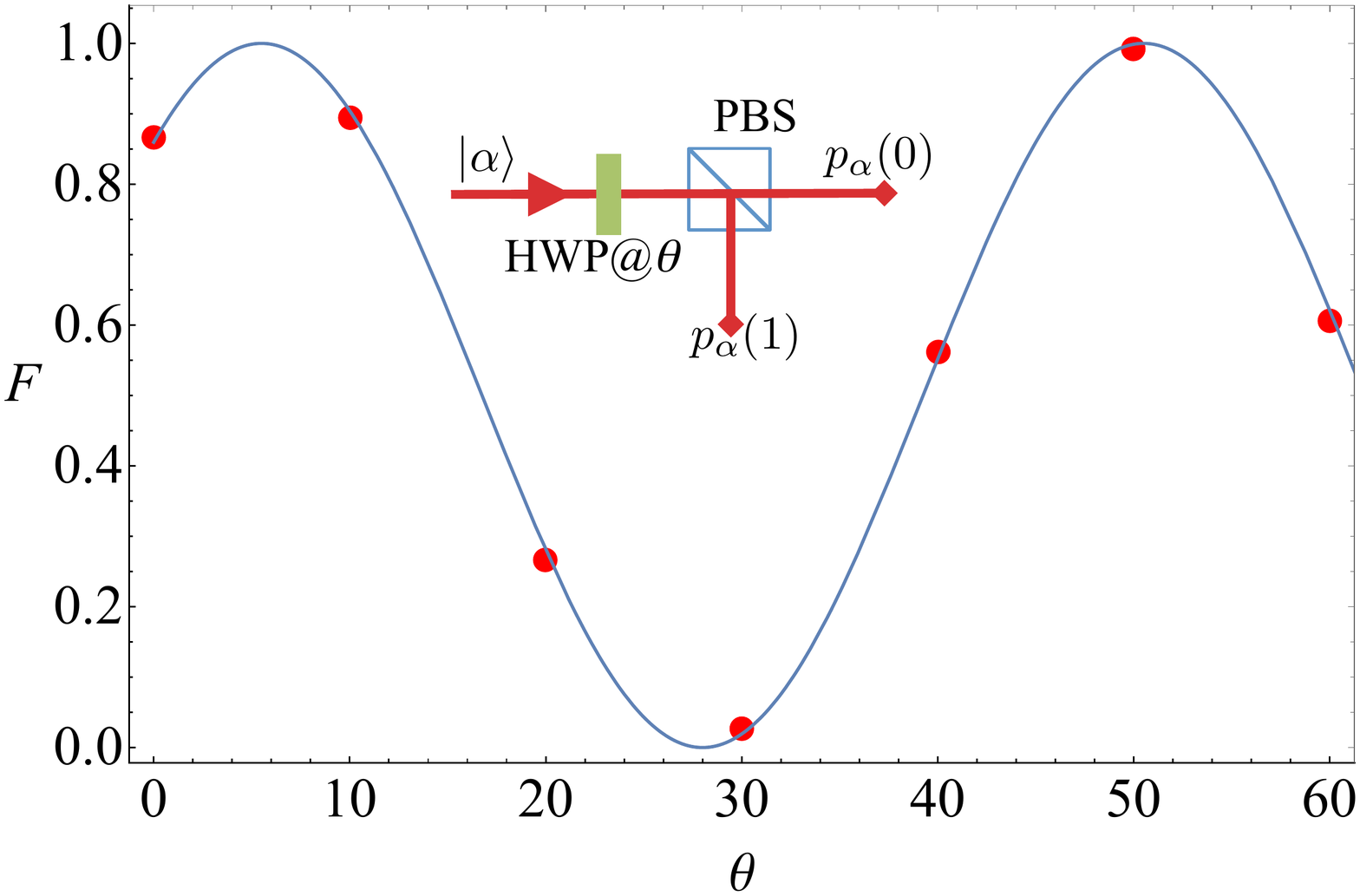}
\caption{Reconstructed Fisher information for two-outcome polarisation measurements. The experimental points correspond to intensity measurements taken on a He:Ne laser with an optical power meter. The solid line indicates the prediction based on a model of the measurement. Inset: experimental setup, consisting of a half waveplate and a polarising beam splitter separating horizonal and vertical components.}
\label{fig:1}
\end{figure}

The search can be effectively restricted to pure states, as these are extremal for $F$~\cite{Pari08}; in this case, this is effectively casted as the minimisation of a function of two parameters. The optimisation procedure needs being carried out cautiously. Small systematic errors in determining the DPFs can originate unphysical maxima, resulting from negative eigenvalues of the measurement matrix. On then needs to introduce the constraint of only looking at states giving positive values for the estimated probabilities. In our example, an optimal measurement is achieved in two conditions: by looking either for a variation of a high contrast between the output probabilities, or for a variation in a balanced condition~\cite{nota1}; the systematic effects explained above favour the second maxima, a result which is consistent with the resilience of these optimal states to small dephasing~\cite{nota1}. Loosing information on the second set of maxima is the price one pays for the simplicity of the method.  The results are summarised in Fig.\ref{fig:1}, where we plot the Fisher information $F$ we obtain as a function of the angle setting $\theta$: our method is able to identify the state that lies closest to the prediction for an ideal Pauli measurement. 



\section{Experimental example 2: single-qubit four-outcome measurement.} 
The next example we consider is acting on a single qubit, performing a $\sigma_z$-measurement half of the times, and $\sigma_x$-measurement the remaining half. Again, we can make a direct parallel with polarimetry, and use it for estimating jointy a phase shift $\phi\sim0$, followed by a rotation $\chi$. The measurement is ideal for small rotations $\chi\sim0$; by ideal here, we do not mean a saturation of the Cram\'er-Rao bound, which is prevented by a Heisenberg-type relation~\cite{massar}, but the saturation of the bound
\begin{equation}
\label{mihai}
\frac{F_{\phi,\phi}}{H_{\phi,\phi}}+\frac{F_{\chi,\chi}}{H_{\chi,\chi}}\leq1,  
\end{equation}
which effectively limits all possible measurements on a single qubit when it comes to two-parameter estimation~\cite{massar,Vidr14}. For a perfect measurement, we expect $\frac{F_{\phi,\phi}}{H_{\phi,\phi}}{=}\frac{F_{\chi,\chi}}{H_{\chi,\chi}}$; the derivation of this bound, first obtained in~\cite{massar},  is presented in Appendix A. However, in our implementation, this condition can be affected by imperfections in the optical components; we use our  DFP method to understand in what measure. The optimisation should not be carried out as a direct optimisation of the bound Eq.\eqref{mihai}, but consider the effective values $F'_{\phi,\phi}=1/(\underline{F}^{-1})_{\phi,\phi}=F_{\phi,\phi}-F^2_{\chi,\phi}/F_{\chi,\chi}$, and $F'_{\chi,\chi}=1/(\underline{F}^{-1})_{\chi,\chi}=F_{\chi,\chi}-F^2_{\chi,\phi}/F_{\phi,\phi}$, {i.e.} the quantities bounding the individual variances for each estimator~\cite{Pari08}. 

\begin{figure}[t]
\includegraphics[viewport = 50 0 900 550, clip, width=\columnwidth]{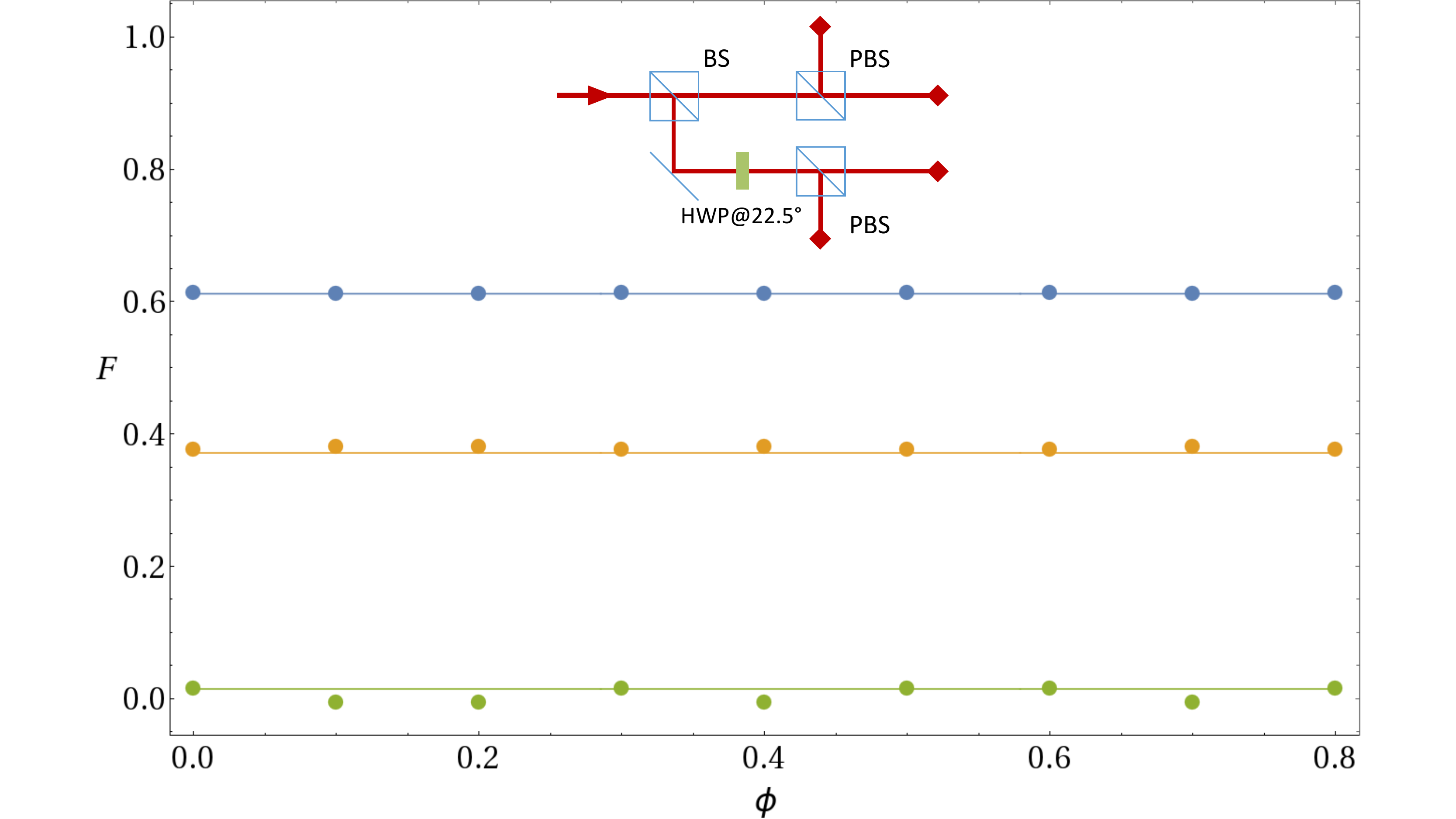}
\caption{Reconstructed Fisher information for a four-outcome polarisation measurement in either the $D/A$ (Pauli $X$) or $H/V$ (Pauli $Z$) basis: $\phi$ is the scanned parameter, while $\chi$ is kept constant at the value 0. The experimental points correspond to intensity measurements taken on a He:Ne laser with an optical power meter: yellow  $F_{\chi,\chi}$, blue: $F_{\phi,\phi}$, green:$F_{\chi,\phi}$. The solid line indicate the prediction based on the reconstructed detector tomography~\cite{Lund09}. Inset: experimental setup, consisting of a non-polarising beam splitter (BS), and two sets of a half waveplate and a polarising beam splitter. The asymmetry in the Fisher information associated to the two parameters arise from the dependence of the splitting ration of BS on the polarisation.} 
\label{fig:2}
\end{figure}

Figure \ref{fig:2} summarises the results of numerical searches of the Fisher information matrix $\underline{F}(\chi,\phi)$, associated to different values of the phase shift $\phi$ for $\chi{=}0$; we also compare them with the predictions obtained by a detector tomography of our apparatus. We observe an uneven splitting of the information between the two parameters, as a result of the experimental imperfections. The direct estimates remain close to those from the tomography, although oftentimes the correlation terms $F_{\chi,\phi}$ present some discrepancies. 

\section{Theoretical example: weak-field homodyne.} Our approach can also be useful for obtaining theoretical predictions when working with measurements whose description involves a complex expression of its POVM. This is the case, notably, for continuous-variable state observed with hybrid detection schemes, mixing elements from photon counting and homodyning~\cite{wfh1,wfh2,wfh25,wfh3,wfh4,wfh5,wfhgaia}. 

In most cases, the most practical choice consists in inspecting the response to coherent states $|\alpha\rangle\langle \alpha|$. By linearity, the response $p_x$ for the outcome $x$ can by expressed in terms of the $P$ representation of the state:
\begin{equation}
\label{1}
p_x=\text{Tr}[\Pi_x\rho]=\int d^2\alpha \, P(\alpha)\, q_x(\alpha).
\end{equation}
This expression is helpful mostly in the classical regime, when the $P$ function has an analytical behaviour, but has limited use for quantum states. This problem can be solved, by using the explicit relation between the $P$-function and the the normally ordered characteristic function:
$\chi_N(\beta)=\text{Tr}[\rho\, e^{i\beta \hat a^\dag} e^{i\beta^* \hat a}]$, giving
\begin{equation}
\label{3}
P(\alpha)=\frac{1}{\pi^2}\int d^2\beta\, \chi_N(\beta) \,e^{-\beta^*\alpha+\beta\alpha^*}
\end{equation}
When using the Fourier relation \eqref{3} in the expression of the probability \eqref{1} for the outcome $x$, we find
\begin{equation}
\label{4}
p_x = \int d^2\beta\, \chi_N(\beta) \,\tilde q_x(\beta),
\end{equation}
where we have introduced the Fourier transform of the DFP 
\begin{equation}
\label{5}
\tilde q_x(\beta)= \frac{1}{\pi^2} \int d^2\alpha\, e^{-\beta^*\alpha+\beta\alpha^*}\, q_x(\alpha).
\end{equation}
This expression is more conveniently cast in terms of the standard symmetric characteristic function which originates the Wigner representation:
\begin{equation}
p_x =  \int d^2\beta\, \chi_S(\beta)\, e^{|\beta|^2/2} \,\tilde q_x(\beta),
\end{equation}
Finally, we can manipulate this expression to make the Wigner function explicitly appear, by using its explicit link to the characteristic function
$\chi_S(\beta)=\int d^2\alpha\, W(\alpha)\, e^{\beta^*\alpha-\alpha^*\beta}$, thus writing
\begin{equation}
\label{8}
p_x = \int d^2\alpha\, W(\alpha) \,\zeta_x(\alpha)
\end{equation}
with 
\begin{equation}
\label{9}
\zeta_x(\alpha)=\int d^2\beta\,e^{\beta^*\alpha-\alpha^*\beta} \, e^{|\beta|^2/2} \,\tilde q_x(\beta).
\end{equation}
Remarkably, our formalism can deal with classical fields through a straightforward correspondence principle, \eqref{1}, as well as with quantum fields, although through a more involved expression, \eqref{8}.

\begin{figure}[t]
\subfloat{\includegraphics[viewport = 0 0 420 350, clip, width =  .5\columnwidth]{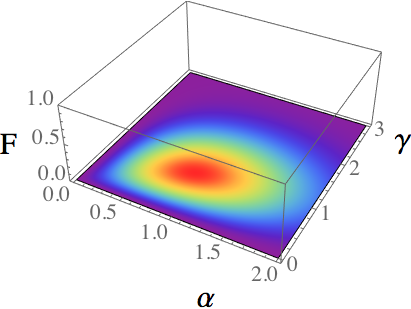}}
\subfloat{\includegraphics[viewport = 0 0 420 330, clip, width =  .5\columnwidth]{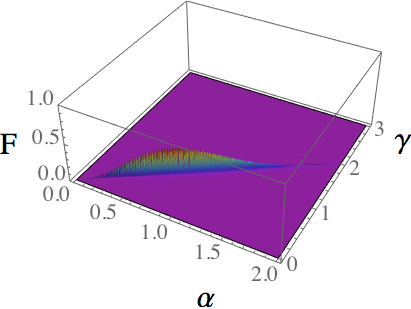}}\\
\subfloat{\includegraphics[viewport = 0 0 420 330, clip, width =  .5\columnwidth]{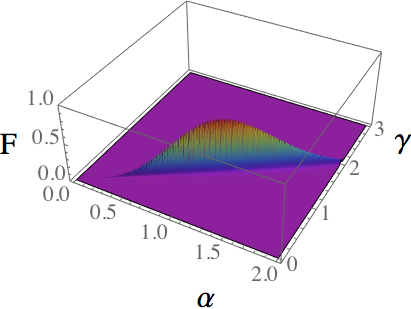}}
\subfloat{\includegraphics[viewport = 0 0 420 330, clip, width =  .5\columnwidth]{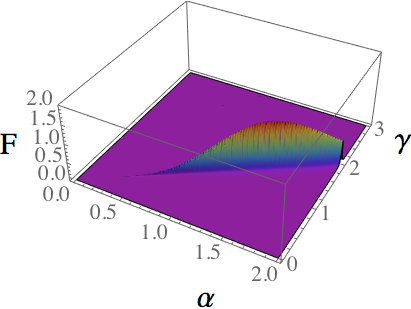}}\\
\subfloat{\includegraphics[viewport = 0 0 420 330, clip, width =  .5\columnwidth]{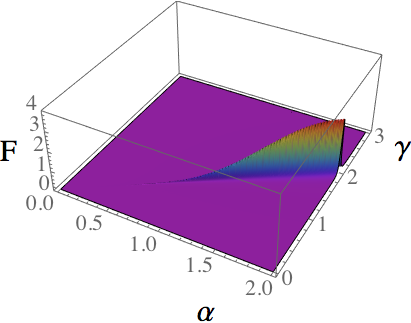}}
\subfloat{\includegraphics[viewport = 0 0 420 330, clip, width =  .5\columnwidth]{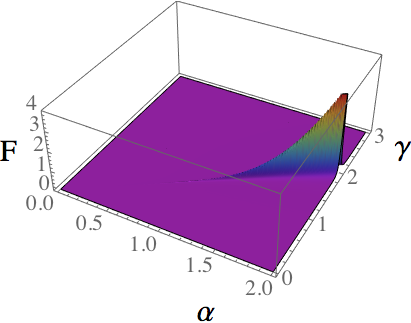}}
\caption{Fisher information associated to different outcomes of a weak-field homodyne for coherent states of real amplitude $\alpha$, for different local oscillators $\gamma$. The outcomes are in the order $(x_1,x_2)=(1,1)$, $(1,0)$, $(0,1)$, $(2,1)$, $(3,1)$, $(4,1)$. The other contributions are negligible in this regime.}
\label{fig:3}
\end{figure}

We now consider the explicit case of the weak-field homodyne~\cite{wfh4,wfh5,wfhgaia}, for which the DFP approach is particularly suited~\cite{hard}. This is the leading example of a hybrid detector: when using this technique a signal is combined on a 50:50 beam splitter with a local oscillator, whose intensity is comparable with that of the signal, hence, in a coherent state $\ket{\gamma}$ at the few-photon level. The two outputs of the beam splitter are then monitored with a photon-counting detector; this is generally implemented dividing light into $N$ bins - either temporal or spatial - each of them measured with a click/no click detector. The corresponding DFPs are written as
\begin{equation}
\begin{aligned}
\label{dpfcv}
& q_x(\alpha) ={N \choose x_1} {N \choose x_2} \sum_{y_1=0}^{x_1}\sum_{y_2=0}^{x_2}
(-1)^{x_1-y_1+x_2-y_2}{x_1 \choose y_1} {x_2 \choose y_2}\\
&\times \exp[-\frac{N-y_1}{2N}|\alpha+\gamma|^2-\frac{N-y_2}{2N}|\alpha-\gamma|^2] 
\end{aligned}
\end{equation}
where $x_1$ ($x_2$) denotes the detection event of the detector on the transmitted (reflected) arm, and we noted $x=\{x_1,x_2\}$.\\
In order to calculate its Fourier transform, we find it convenient to consider separately each term in the sum, whose Fourier transform, according to the general expression \eqref{9} takes the form:
\begin{equation}
\tilde q_x(\beta)\propto\frac{\sigma^2}{\pi} e^{-\sigma^2\left(|\beta|^2-|\tilde\gamma|^2+\beta^*\tilde\gamma-\beta\tilde\gamma^*\right)} e^{-\frac{|\gamma|^2}{\sigma^2}}
\end{equation}
with the shorthand notation $\tilde \gamma{=}\frac{y_2-y_1}{2N}\gamma$, and $\sigma^{-2}{=}\frac{2N-y_1-y_2}{2N}$. Finally, the convolution \eqref{9}, gives the expression 
\begin{equation}
\zeta_x(\alpha)\propto \frac{\sigma^2}{\sigma^2-\frac{1}{2}}e^{-\frac{|\alpha|^2+\sigma^2|\tilde \gamma|^2/2}{\sigma^2-{1}/{2}}}e^{\frac{\sigma^2}{\sigma^2-1/2} \alpha^*\tilde\gamma+\alpha\tilde\gamma^*} e^{-\frac{|\gamma|^2}{\sigma^2}}.
\end{equation}
from which one can then calculate the Fisher information associated to an arbitrary state via \eqref{8}. The full derivation of the expression above is detailed in Appendix B.

\begin{figure}[b]
\includegraphics[viewport = 0 0 560 380, clip, width= \columnwidth]{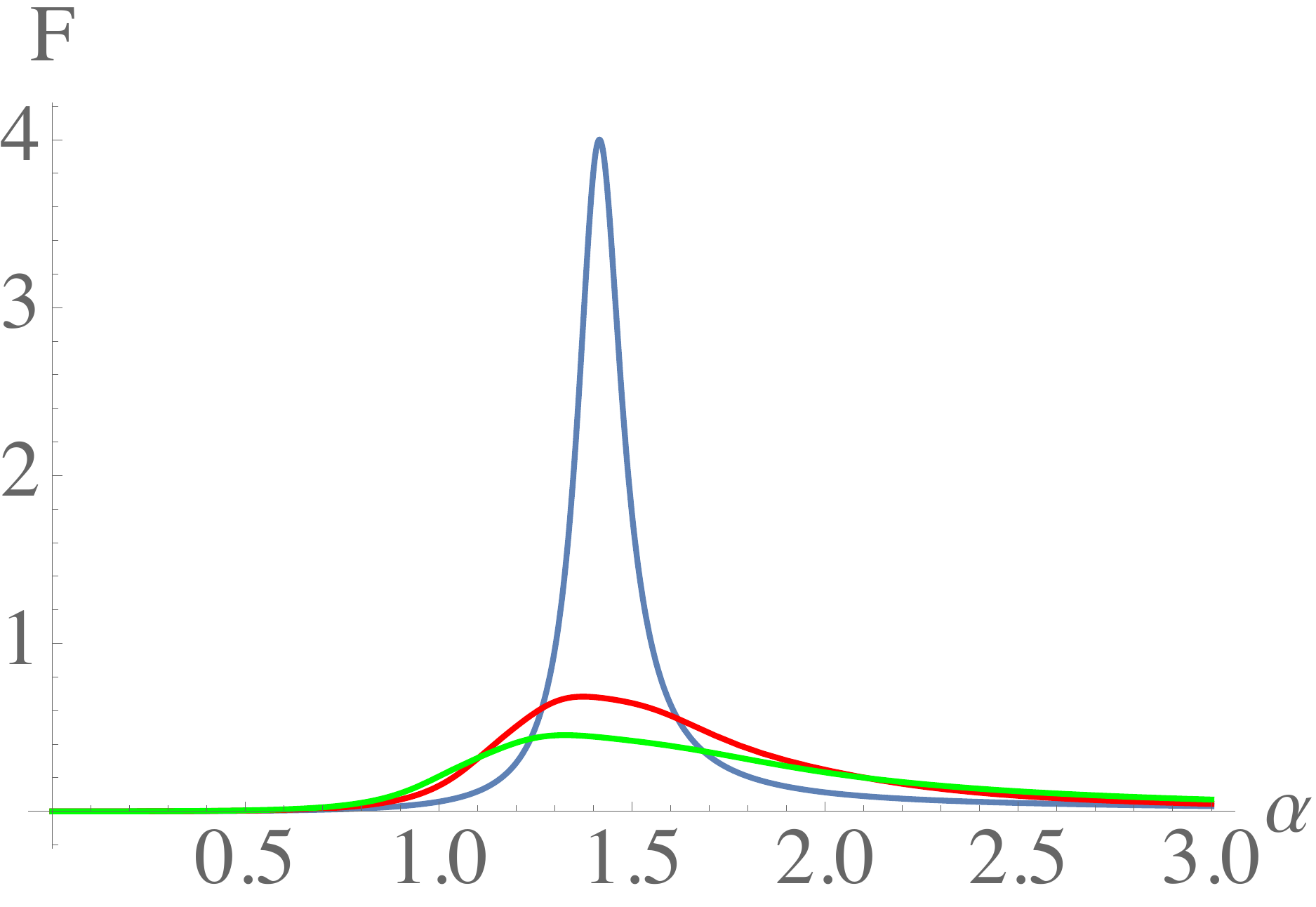}
\caption{Fisher information for a phase measurement ($\phi=0.1$) with squeezed probe states of given energy $\alpha^2$ divided partly into the displacement with $r_d=\alpha_0^2/\alpha^2$, and the rest into squeezing in the $P$ direction. We report $r_d=1$ in blue, $r_d=0.95$ in red, and $r_d=0.9$ in green. }
\label{fig:s1}
\end{figure}

The Fisher information on a small phase shift $\phi$ attached to each outcome \eqref{dpfcv} $F^{(x)}=\left(\partial_\phi q_{x}(\alpha)\right)^2/q_{x}(\alpha)$ is shown in Fig.\ref{fig:3} for the case $N=4$. Inspection of these curves reveals that, for a given local oscillator $\gamma$, the Fisher information is concentrated in a few detection events, and it is sharply peaked around $\alpha=\gamma$. Furthermore, being the local oscillator in a coherent state with no correlations to the probe, fluctuations in the photon number can not be suppressed below the Poissonian noise. Therefore, coherent states are optimal probes for this detection scheme, and, despite the close resemblance to ordinary strong-field homodyne, there is no advantage in utilising squeezing. 

In order to confirm these observations, we adopt the DFP formalism to compute the Fisher information associated to squeezed states when estimating a small phase, and compare it to the benchmark provided by coherent states. We do so at a given energy, remembering that for a displaced squeezed vacuum $W(\alpha){=}\frac{2}{\pi} \exp{\left[-2s^2(\alpha_x-\alpha_0)^2-2\alpha_p^2/s^2\right]}$, where $\alpha=\alpha_x+i\alpha_p$, and $s<1$ ($s>0$) quantifies the squeezing in the $P$ ($X$) quadrature; the average photon number in this state is $\alpha_0^2+\frac{(s^2-1)^2}{4s^2}$. The typical behaviour is illustrated in Fig.\ref{fig:s1}, where we show the Fisher information at a given total energy $\alpha^2$, split differently between the displacement and the squeezing: the addition of squeezing rapidly degrades the informational content of the measurement, as we were expecting from the analysis of the DPFs.

\section{Conclusions.} We have introduced a new approach to the optimisation of probe states for parameter estimation in the presence of a detector whose exact response is unknown. It offers a computational advantage with respect the standard approach based on detector tomography, although, in the absence of regularisation, suitable filters have to be applied. 

In addition, we have shown how this formalism also provides an agile theoretical description of continuous-variable measurement devices, and illustrated its use for weak-field homodyne. This has also potential for applications for experimental characterisation, but it should be adapted to the availability of a limited number of states to be used~\cite{Coop14}.

\section*{Acknowledgements.} We thank Paolo Aloe for technical support. This project has received funding from the European UnionÕs Horizon 2020 research and innovation programme under grant agreement No 665148.  M.B. has been supported by a Rita Levi-Montalcini fellowship of MIUR. M.G.G. acknowledges support from EPSRC through Grant No. EP/K026267/1.

\section*{Appendix A: two-phase estimation}

We consider an experiment in which the two parameters to be estimated are a phase shift $\phi$ around the $Z$ axis of the Poincar\'e-Bloch sphere of a single qubit, and, at the same time, a phase shift $\chi$ around the $X$ axis. The two operations are described by:
\begin{equation}
\begin{aligned}
U&=\left(
\begin{array}{cc}
e^{-i \phi/2}&0\\
0&e^{i \phi/2}
\end{array}\right),\\
V&=\left(
\begin{array}{cc}
\cos( \chi/2)&-i \sin( \chi/2)\\
-i \sin( \chi/2)&\cos( \chi/2)
\end{array}
\right).
\end{aligned}
\end{equation}
These operators do not commute, so we have to consider the two possible permutations separately. First, let us consider the product $V\cdot U$ applied to the state $\ket{\psi}=\frac{1}{\sqrt{2}}(1,i)$, dictated by symmetry. Its quantum Fisher information matrix ${\underline H}$ is found to be
\begin{equation}
\begin{aligned}
{\underline H}&=\left(
\begin{array}{cc}
1&0\\
0&\cos^2(\phi)
\end{array}\right).
\end{aligned}
\end{equation}
For the estimation, we consider explicitly a POVM which projects along either the $Z$ or the $X$ axis with equal probability; explicit calculations give  
\begin{equation}
\begin{aligned}
F'_{\phi,\phi}&=\frac{1}{2},\\
F'_{\chi,\chi}&=\frac{\cos^2(\phi) \cos^2(\chi)}{2-2\cos(2\phi)\sin^2(\chi)}.
\end{aligned}
\end{equation}
These saturate the bound $\frac{F'_{\phi,\phi}}{H_{\phi,\phi}}+\frac{F'_{\chi,\chi}}{H_{\chi,\chi}}\leq1$ for $\chi=0$. In this case, one can achieve 
$F'_{\chi,\chi}=1/2$ by an adaptive estimation of the phase $\phi$. In the second instance  $U\cdot V$, we obtain for the quantum Fisher information matrix
\begin{equation}
\begin{aligned}
{\underline H}&=\left(
\begin{array}{cc}
\cos^2(\chi)&0\\
0&1
\end{array}\right),
\end{aligned}
\end{equation}
If one considers the same POVM as above:
\begin{equation}
\begin{aligned}
F'_{\phi,\phi}&=\frac{\cos^2(\phi) \cos^2(\chi)}{2-2\cos(2\chi)\sin^2(\phi)},\\
F'_{\chi,\chi}&=\frac{1}{2}.
\end{aligned}
\end{equation}
Similar considerations as for the first case are valid: optimal estimation occurs for $\phi=0$, with the possibility of achieving the highest information by means of an adaptive protocol for $\chi$.

\section*{Appendix B: data fitting pattern for the weak-field homodyne}

Here we detail the derivation of the formulae describing the data pattern functions associated to a generic cv measurement device, and, in particular, to the weak-field homodyne.
 
The data fitting pattern method consists in describing a detector by inspecting its response to coherent states $|\alpha\rangle\langle \alpha|$.
By linearity, the response $g_x$ for the outcome $x$ is given by:
\begin{equation}
\label{1}
p_x=\text{Tr}[\Pi_x\rho]=\int d^2\alpha\, P(\alpha)\, \text{Tr}[\Pi_x|\alpha\rangle\langle \alpha|]=\int d^2\alpha \, P(\alpha)\, q_x(\alpha).
\end{equation}
The main problem of this expression is in the bad behaviour of the $P$-function. This can be solved, by using the relation of the $P$-function
to the normally ordered characteristic function:
\begin{equation}
\chi_N(\beta)=\text{Tr}[\rho\, e^{i\beta \hat a^\dag} e^{i\beta^* \hat a}]
\end{equation}
giving
\begin{equation}
\label{3}
P(\alpha)=\frac{1}{\pi^2}\int d^2\beta\, \chi_N(\beta) \,e^{-\beta^*\alpha+\beta\alpha^*}
\end{equation}
When using the Fourier relation \eqref{3} in the expression of the probability \eqref{1} for the outcome $x$, we find
\begin{equation}
\label{4}
\begin{aligned}
p_x &= \frac{1}{\pi^2} \int d^2\alpha \,\int d^2\beta\, \chi_N(\beta) \,e^{-\beta^*\alpha+\beta\alpha^*}\, q_x(\alpha)\\
&= \frac{1}{\pi^2} \int d^2\beta\, \chi_N(\beta) \,\int d^2\alpha\, e^{-\beta^*\alpha+\beta\alpha^*}\, q_x(\alpha)\\
&=  \int d^2\beta\, \chi_N(\beta) \,\tilde q_x(\beta),
\end{aligned}
\end{equation}
where we have introduced the Fourier transform of the DFP 
\begin{equation}
\label{5}
\tilde q_x(\beta)= \frac{1}{\pi^2} \int d^2\alpha\, e^{-\beta^*\alpha+\beta\alpha^*}\, q_x(\alpha).
\end{equation}
This expression is more conveniently cast in terms of the standard symmetric characteristic function which originates the Wigner representation:
\begin{equation}
p_x =  \int d^2\beta\, \chi_S(\beta)\, e^{|\beta|^2/2} \,\tilde q_x(\beta),
\end{equation}
We can manipulate this expression to make the Wigner function explicitly appear, by recalling its explicit link to the characteristic function
\begin{equation}
\chi_S(\beta)=\int d^2\alpha\, W(\alpha)\, e^{\beta^*\alpha-\alpha^*\beta}.
\end{equation}
We then get
\begin{equation}
\begin{aligned}
\label{8}
p_x &=  \int d^2\beta\, \int d^2\alpha\, W(\alpha)\, e^{\beta^*\alpha-\alpha^*\beta} \, e^{|\beta|^2/2} \,\tilde q_x(\beta)\\
&= \int d^2\alpha\, W(\alpha) \,\zeta_x(\alpha)
\end{aligned}
\end{equation}
with 
\begin{equation}
\label{9}
\zeta_x(\alpha)=\int d^2\beta\,e^{\beta^*\alpha-\alpha^*\beta} \, e^{|\beta|^2/2} \,\tilde q_x(\beta).
\end{equation}
We now consider the explicit case of weak-field homodyne, with a local oscillator $\gamma$ and $N$ detection bins. The corresponding DFPs are written as
\begin{equation}
\begin{aligned}
& q_x(\alpha) ={N \choose x_1} {N \choose x_2} \sum_{y_1=0}^{x_1}\sum_{y_2=0}^{x_2}(-1)^{x_1-y_1+x_2-y_2}{x_1 \choose y_1} {x_2 \choose y_2}\\
&\times \exp[-\frac{N-y_1}{2N}|\alpha+\gamma|^2-\frac{N-y_2}{2N}|\alpha-\gamma|^2] 
\end{aligned}
\end{equation}
where $x_1$ ($x_2$) denotes the detection event of the detector on the transmitted (reflected) arm, and we have denoted $x=\{x_1,x_2\}$.
In order to calculate its Fourier transform, we consider separately each term in the sum:
\begin{widetext}
\begin{equation}
q_x(\alpha)\propto\exp[{-\beta^*\alpha+\beta\alpha^*}]\exp[-\frac{N-y_1}{2N}|\alpha+\gamma|^2-\frac{N-y_2}{2N}|\alpha-\gamma|^2] 
\end{equation}
If we now expand the moduli, we obtain
\begin{equation}
\begin{aligned}
q_x(\alpha)&\propto\exp[{-\beta^*\alpha+\beta\alpha^*}]\exp[-\frac{N-y_1}{2N}\left(|\alpha|^2+|\gamma|^2+\alpha^*\gamma+\alpha\gamma^*)\right )]\\
&\times \exp[-\frac{N-y_2}{2N}\left(|\alpha|^2+|\gamma|^2-\alpha^*\gamma-\alpha\gamma^*)\right )]. 
\end{aligned}
\end{equation}

We can then group the exponentials more sensibly 
\begin{equation}
\label{13}
\begin{aligned}
&\exp\left[{\alpha^*\left(\beta+\gamma\left({\frac{N-y_2}{2N}-\frac{N-y_1}{2N}}\right)\right)-\alpha\left(\beta^*-\gamma^*\left({\frac{N-y_2}{2N}-\frac{N-y_1}{2N}}\right)\right)}\right]\\
&\times \exp\left[-|\alpha|^2\left(\frac{N-y_2}{2N}+\frac{N-y_1}{2N}\right)\right]\exp\left[-|\gamma|^2\left(\frac{N-y_2}{2N}+\frac{N-y_1}{2N}\right)\right]\\
&=\exp[\alpha^*(\beta+\tilde\gamma)-\alpha(\beta^*-\tilde\gamma^*)]\exp\left[-\frac{|\alpha|^2}{\sigma^2}\right]\exp\left[-\frac{|\gamma|^2}{\sigma^2}\right]
\end{aligned}
\end{equation}
with the obvious meaning of the symbols. We find that the Fourier transform \eqref{5} has the form of a sum of terms in the form:
\begin{equation}
\tilde q_x(\alpha)\propto{\pi \sigma^2} e^{-\sigma^2\left(|\beta|^2-|\tilde\gamma|^2+\beta^*\tilde\gamma-\beta\tilde\gamma^*\right)} 
\end{equation}
The convolution \eqref{9} is in a similar form, with each term given by
\begin{equation}
\begin{aligned}
\zeta_x(\alpha)&\propto{\pi \sigma^2}e^{-\frac{|\gamma|^2}{\sigma^2}} \int d^2\beta\,e^{\beta^*\alpha-\alpha^*\beta} \, e^{|\beta|^2/2} \,e^{-\sigma^2\left(|\beta|^2-|\tilde\gamma|^2+\beta^*\tilde\gamma-\beta\tilde\gamma^*\right)}\\
&={\pi^2} \frac{\sigma^2}{\sigma^2-\frac{1}{2}}e^{-\frac{|\alpha|^2+\sigma^2|\tilde \gamma|^2/2}{\sigma^2-{1}/{2}}}e^{\frac{\sigma^2}{\sigma^2-1/2} \alpha^*\tilde\gamma+\alpha\tilde\gamma^*} e^{-\frac{|\gamma|^2}{\sigma^2}}.
\end{aligned}
\end{equation}
\end{widetext}


\begin{thebibliography}{99}
\bibitem{Giov04} V. Giovannetti, S. Lloyd, and L. Maccone, Science 306, 1330 (2004).
\bibitem{Giov06} V. Giovannetti, S. Lloyd, and L. Maccone, Phys. Rev. Lett. 96, 010401 (2006).
\bibitem{Pari08}  M.G.A. Paris, 	Int. J. Quant. Inf. 7, 125 (2009).
\bibitem{Kalt08} R. Kaltenbaek, J. Lavoie, D.N. Biggerstaff, and K.J. Resch, Nature Phys 4, 864 (2008).
\bibitem{Dixo09} P.B. Dixon, D.J. Starling, A.N. Jordan, and J.C. Howell, Phys. Rev. Lett. 102, 173601 (2009).
\bibitem{Host08} O. Hosten, and P.G. Kwiat, Science 319, 787 (2008).
\bibitem{Pfei11} M. Pfeister, and P. Fisher, Opt. Exp. 19, 16508 (2011).
\bibitem{Dorn09} U. Dorner, {\it et al.}, Phys. Rev. Lett. 102, 040403 (2009).
\bibitem{Datt11} A. Datta, L. Zhang, N. Thomas-Peter, U. Dorner, B.J. Smith, and I.A. Walmsley, Phys. Rev. A 83, 063836 (2011). 
\bibitem{Knys11} S. Knysh, V.N. Smelyanskiy, and G.A. Durkin,  Phys. Rev. A 83, 021804(R) (2011).
\bibitem{Geno11} M.G. Genoni, S. Olivares, and M.G.A. Paris, Phys. Rev. Lett. 106, 153603 (2011).
\bibitem{Geno12} M.G. Genoni et al., Phys. Rev. A 85 (4), 043817 (2012).
\bibitem{Lund09} J.S. Lundeen, {\it et al.}, Nature Phys. 5, 27 (2009).
\bibitem{Hels76} C. W. Helstrom, {\it Quantum Detection and Estimation Theory}, Mathematics in Science and Engineering Vol. 123 (Academic, New York, 1976).
\bibitem{Reac10} J. \v Reh\'a\v cek, D. Mogilevtsev, and Z. Hradil, Phys. Rev. Lett. 105, 010402 (2010).
\bibitem{Coop14} M. Cooper, M. Karpi\'nski, and B.J. Smith, Nature Comms. 4, 4332 (2014).
\bibitem{Brid12} G. Brida, {\it et al.}, New J. Phys. 14, 085001 (2012). 
\bibitem{Zhan12} L. Zhang, {\it et al.}, Nature Photon, 6, 364 (2012).
\bibitem{Gran15} S. Grandi, A. Zavatta, M. Bellini, and M.G.A. Paris, arXiv:1505.03297 (2015).
\bibitem{silb} G. Harder et al., Phys. Rev. A 90, 042105 (2014). 
\bibitem{nota1}  D. Brivio {\it et al}, Phys. Rev. A. 81, 012305 (2010). 
\bibitem{massar} S. Massar, and R. Gill, Phys. Rev. A. 61, 042312 (2000).  
\bibitem{Vidr14} M.D. Vidrighin {\it et al.}, Nature Comms 5, 3532 (2014).
\bibitem{wfh1} A. Kuzmich, I. A. Walmsley, and L. Mandel,  Phys. Rev. Lett. 85, 1349 (2000).
\bibitem{wfh2} K.J. Resch, J.S. Lundeen, and A.M. Steinberg, Phys. Rev. Lett. 88, 113601 (2002).
\bibitem{wfh25} B. Hessmo, P. Usachev, H. Heydari, and G. Bj\"ork, Phys. Rev. Lett. 92, 180401 (2004).
\bibitem{wfh3} G. Puentes, G. et al. Phys. Rev. Lett. 102, 080404 (2009).
\bibitem{wfh4} K. Laiho, K.N. Cassemiro, D. Gross, and C. Silberhorn, Phys. Rev. Lett. 105, 253603 (2010).
\bibitem{wfh5} L. Zhang, et al. Nat. Photonics 6, 364Ð368 (2012).
\bibitem{wfhgaia} G. Donati et al. Nat. Comms 5, 5584 (2014).
\bibitem{hard} G. Harder et al., in preparation.


\end{thebibliography}
\end{document}